\documentclass[pre,twocolumn,showpacs]{revtex4}

\usepackage{graphicx}

\begin{document}

\title{Statistical fluctuations of the parametric derivative of the
transmission and reflection coefficients in absorbing chaotic
cavities}

\author{M. Mart\'{\i}nez-Mares}

\affiliation{Departamento de F\'{\i}sica, Universidad Aut\'onoma
Metropolitana-Iztapalapa, Av. San Rafael Atlixco 186, Col. Vicentina,
09340 M\'exico D. F., M\'exico}

\date{\today}


\begin{abstract}
Motivated by recent theoretical and experimental works, we study the statistical
fluctuations of the parametric derivative of the transmission $T$ and reflection
$R$ coefficients, $\partial T/\partial X$ and $\partial R/\partial X$
respectively, in ballistic chaotic cavities in the presence of absorption.
Analytical results for the variance of $\partial T/\partial X$ and
$\partial R/\partial X$, with and without time-reversal symmetry, are obtained
for asymmetric and left-right symmetric cavities. These results are valid
for an arbitrary number of channels for strong absorption strength, in complete
agreement with the results found in the literature in the absence of absorption.
A simple extrapolation to any absorption strength is qualitatively correct.
\end{abstract}

\pacs{05.45.Mt, 03.65.Nk, 73.23.-b}

\maketitle


\section{Introduction}
\label{sec:intro}

In chaotic and weakly disordered quantum systems which are not self-averaging,
phase coherence gives rise to sample-to-sample fluctuations in most transport
properties with respect to a small perturbation in the incident energy, an
applied magnetic field or the shape of the system. Those fluctuations are
universal \cite{Beenakker1997,Alhassid} and depend only on the symmetry
properties, such as the presence or absence of time reversal invariance (TRI),
and spatial symmetry \cite{MQP-1994,Gopar1996,Baranger1996,Martinez2000}. An
statistical analysis is well described by random matrix theory (RMT) \cite{Metha}.

The parametric dependence of the conductance has been studied experimentally by
considering ballistic quantum dots connected to electron reservoirs by ballistic
points contacts with few propagating modes \cite{Marcus,Chang,Chan,Keller,Huibers}.
RMT predictions can also be verified in wave scattering experimental systems, such
as microwave cavities \cite{Richter,Stoekmann}, acoustic resonators \cite{Schaadt},
or elastic media \cite{Mori}, where the external parameters are easy to control.
However, absorption is always present in these experiments and its influence on the
universal transport properties is rather dramatic \cite{Doron1990}; therefore, many
theoretical and experimental works have been devoted to the effect of absorption on
the transmission $T$ and reflection $R$ coefficients of the cavity \cite{Doron1990,
Brouwer1996,Kogan2000,Beenakker2001,Schanze2001,Savin2003,Mendez-Sanchez2003}. The
derivative of those coefficients with respect to the external parameter has not
been considered in the presence of absorption. A parametric derivative is very
important in the characterization of mesoscopic systems with a chaotic classical
limit \cite{Brouwer1997,mc}, since it is analogous to the level velocity
\cite{Simons,Fyodorov1,Taniguchi,Fyodorov2}.

Motivated by recent experiments in microwave cavities \cite{Mendez-Sanchez2003,
Schanze2001}, in the present paper we study the statistical fluctuations of the
parametric derivative of $T$ and $R$ with respect to an external parameter $X$,
$\partial T/\partial X$ and $\partial R/\partial X$, in the presence of absorption.
We consider a chaotic cavity connected to two waveguides with an arbitrary number
of channels, with and without TRI, and we address both asymmetric and left-right
(LR) symmetric cavities. As an external parameter we will take shape deformations.
The purpose of this work is three fold: first, the calculations here presented help
to understand the distribution of the energy derivative of $T$ in the presence
of absorption; in fact, we now present a complete theoretical derivation of some of
the results used in Ref. \onlinecite{Schanze2003}. Second, they also can serve to
motivate the experimental analysis of the distribution of the derivative of $T$ but with respect to shape deformations, where the results of the present paper can be applied. That is the case of Ref. \onlinecite{Schanze2003} where, in order to improve statistics, the shape is modified by varying one lenght of the resonator used in the experiments. 
Finally, in a similar way, the experimental situation of Ref.
\onlinecite{Mendez-Sanchez2003} can be used as well to study energy and shape deformation derivatives of $R$.

The results presented here are valid for strong absorption. However, they
reproduce those existing in the literature for the distribution of
$\partial T/\partial X$ at zero absorption intensity \cite{Brouwer1997,mc}. In the
absence of absorption the distribution of the parametric conductance derivative was
calculated analytically by Brouwer {\it et al} \cite{Brouwer1997} for an
asymmetric quantum dot with two single-mode point contacts. The
$\partial T/\partial X$-distribution has algebraic tails and in the absence
(presence) of TRI it shows a cusp (divergence) at zero derivative; the second
moment is finite (infinite). The reflection symmetric case was considered in Ref.
\onlinecite{mc}. There, the distribution of $\partial T/\partial X$ diverges
logarithmically at zero derivative, it has algebraic tails with an exponent which is different to that of the asymmetric case.

The paper is organized as follows. In Sect. \ref{sec:theory} we present the
main formal elements used throughout the paper, such as the scattering matrix
$S$ and its parametric derivative in the presence of absorption. Sect.
\ref{subsec:theory-asymm} is dedicated to asymmetric cavities. The Poisson
kernel for $S$ and its application to chaotic scattering in the presence of
absorption is presented by means of a phenomenological model; the parametric
derivative of $S$ is defined in terms of a Wigner time-delay matrix whose
eigenvalues are the proper time-delays, the inverse of them being distributed
according to the Laguerre ensemble. The general structure for $S$ and its
parametric derivative for cavities with LR symmetry is introduced in Sect.
\ref{subsec:theory-symm}. The mean and variance of the parametric velocities
for $T$ and $R$, as well as the correlator between the channel-channel
transmission and reflection coefficients are calculated in Sect.
\ref{subsec:asymm-beta1} in the presence of TRI, and in Sect.
\ref{subsec:asymm-beta2} in its absence. Sect. \ref{sec:symm-fluc} is dedicated
to LR-symmetric cavities where we calculate the variances of parametric
velocities in the presence (absence) of TRI. Finally, a summary of the results as
well as the conclusions are presented in Sect. \ref{sec:conclusions}.


\section{The $S$ matrix and its parametric derivative}
\label{sec:theory}

\subsection{Chaotic scattering by asymmetric cavities in the presence
of absorption}
\label{subsec:theory-asymm}

\begin{figure}
\includegraphics[width=5.0cm]{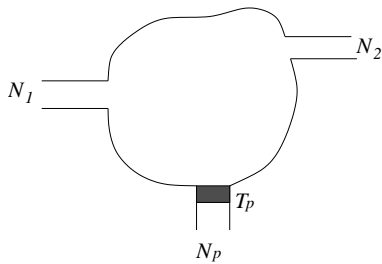}
\caption{A ballistic chaotic cavity connected to two leads with $N_1$,
$N_2$ channels. $N_p$ equivalent ``parasitic'' channels are attached to
the cavity by tunnel barriers with transmission $T_p$ \cite{Lewenkopf1992}. The
absorption strength is given $\gamma_p=N_pT_p$ in the limit $N_p\rightarrow\infty$,
$T_p\rightarrow 0$ keeping the product constant
\cite{Brouwer1997}.}
\label{fig:cavity}
\end{figure}

The scattering problem of a ballistic cavity connected to two waveguides, each
supporting $N_1$, $N_2$ transverse propagating modes (see Fig.
\ref{fig:cavity}), can be described by the scattering matrix $S$ which, in the
stationary case, relates the outgoing to the incoming wave amplitudes
\cite{Newton1982}.

The absorption in the cavity is modeled attaching $N_p$ equivalent non
transmitting or ``parasitic'' channels to the cavity by means of a tunnel barrier
with transmission $T_p$ for each one \cite{Brouwer1997,Lewenkopf1992}. The $S$
matrix is $N$ dimensional ($N=N_1+N_2+N_p$) with a structure given by
\begin{eqnarray} \label{S-abs}
S = \left(\begin{array}{ccc}
s_{11} & s_{12} & s_{1p} \\
s_{21} & s_{22} & s_{2p} \\
s_{p1} & s_{p2} & s_{pp}
\end{array}\right)
\equiv
\left(\begin{array}{cc}
\widetilde{S} & \begin{array}{r}
		s_{1p} \\
		s_{2p}
		\end{array} \\
\begin{array}{cc}
s_{p1} & s_{p2}
\end{array} & s_{pp}
\end{array}
\right),
\end{eqnarray}
where the set of indices $\{1\}$, $\{2\}$, $\{p\}$ label the $N_1$, $N_2$, $N_p$
channels. Here, the submatrix $\widetilde{S}$ of dimension $N_1+N_2$ describes
the scattering problem of the absorbing system. The absorption can be quantified
by the parameter $\gamma_p=N_pT_p$ in the limit $N_p\rightarrow\infty$,
$T_p\rightarrow 0$ while keeping the product constant \cite{Brouwer1997}.

$T$ and $R$ are obtained from $S$, actually only from $\widetilde{S}$, as follows
\begin{equation} \label{TRAs}
T = \sum_{a\in 1} \sum_{b\in 2} \left| S_{ab} \right|^2
\quad \text{and} \quad
R = \sum_{a,b\in 1} \left| S_{ab} \right|^2 .
\end{equation}

In our case only two of the three basic symmetry classes in the Dyson's scheme
\cite{Dyson1962} are relevant . We assume that $S$ satisfy flux conservation by
the restriction
\begin{equation} \label{unitary}
SS^{\dagger} = \openone_N \, ,
\end{equation}
where $\openone_N$ stands for the unit matrix of dimension $N$. This case is
called ``unitary'' and it is designated as $\beta=2$. In addition, in the
presence of time reversal invariance $S$ is symmetric,
\begin{equation} \label{unitary-symmetric}
S=S^T \, .
\end{equation}
This is the ``orthogonal'' case, designated as $\beta=1$. Note that the $N_p$ channels are normal scattering channels for the matrix $S$, while they are absorbing channels for the matrix 
$\widetilde{S}$, which is a subunitary one and describes the physical system; it represents the scattering
matrix of the absorbing system where the flux is not conserved.

For systems with a chaotic classical limit, most transport properties are
sample specific and a statistical analysis of the quantum-mechanical problem is
needed. That study is performed by the construction of ensembles of physical
systems, described mathematically by ensembles of $S$ matrices distributed
according to a probability law. The starting point is a uniform distribution
where $S$ is a member of one of the {\it circular ensembles}: circular unitary
(orthogonal) ensemble, CUE (COE), for $\beta=2$ ($\beta=1$) \cite{Mello1995}.

In the presence of direct processes, the information-theoretic approach
of Refs. \onlinecite{Mello1985}, \onlinecite{Friedman1985} leads to an $S$
matrix distributed according to Poisson's kernel \cite{Mello1999}
\begin{equation} \label{Pkernel}
P_K^{(\beta)}(S) = C
\frac{ \left[ \det \left( \openone_N - \langle S \rangle \langle S
\rangle^{\dagger} \right) \right]^{(\beta N+2-\beta)/2} }
{\left| \det \left( \openone_N - S \langle S \rangle^{\dagger} \right)
\right|^{\beta N+2-\beta} } ,
\end{equation}
where $\langle S\rangle$ is the ensemble averaged $S$ matrix.

A useful model to construct the Poisson ensemble consist of a cavity connected
to leads by tunnel barriers \cite{Brouwer1995-7}. In the case we are concerned with,
where only the fictitious waveguide contains a tunnel barrier, the averaged $S$
matrix can be written as
\begin{equation}
\langle S \rangle = \left(
\begin{array}{ccc}
0_{N_1} & 0 & 0 \\  0 & 0_{N_2} & 0 \\ 0 & 0 & \sqrt{1-T_p}
\openone_{N_p}
\end{array}
\right) \, .
\end{equation}
As before, $\openone_n$ stands for the unit matrix of dimensions $n$ and
$0_n$ for the $n$-dimensional null matrix.

In what follows we restrict ourselves to the case where $T_p=1$, i.e.
$P_K^{(\beta)}(S)$ is just a constant and the $S$ matrix is uniformly distributed.
In this case, we are restricted to a strong absorption situation, where the
parameter $\gamma_p$ takes only integer values ($\gamma_p=N_p$). Also, the results
here presented are valid for no absorption ($N_p=0$), and a simple extrapolation to
non integer values of $\gamma_p$ is qualitatively correct, as will show later on.

If the coupling to the fictitious waveguide is perfect, we can use the well known definition of the parametric derivative of $S$. The derivative of $S$ with respect to the energy of incidence $E$ can be defined in terms of a symmetrized form of the Wigner-Smith time delay matrix \cite{WignerSmith}, whose eigenvalues are identical
among them \cite{BrouwerPRL1997}. In dimensionless units we have
\begin{equation}\label{dES}
\frac{\partial S}{\partial \varepsilon} = i S^{1/2}\, Q_{\varepsilon}\, S^{1/2},
\end{equation}
where we have defined $\varepsilon=2\pi E/\Delta$ with $\Delta$ the mean level
spacing, $Q_{\varepsilon}$ is an $N\times N$ Hermitian matrix for $\beta=2$, real
symmetric for $\beta=1$. The eigenvalues of $Q_{\varepsilon}$ are $\tau_H^{-1}$
times the proper delay times, where $\tau_H=2\pi\hbar/\Delta$ is the Heisenberg
time. In an analogous way, the derivative of $S$ with respect to an external
parameter $X$ is defined as \cite{BrouwerPRL1997}
\begin{equation}\label{dXS}
\frac{\partial S}{\partial x} = i S^{1/2} \, Q_x \, S^{1/2},
\end{equation}
where we have also defined a dimensionless parameter $x=X/X_c$ with $X_c$ a
typical scale for $X$, and $Q_x$ is an $N\times N$ Hermitian matrix, real
symmetric in the presence of time-reversal symmetry.

For classically chaotic cavities the joint distribution of $S$, $Q_{\varepsilon}$
and $Q_x$ is given by \cite{BrouwerPRL1997}
\begin{eqnarray}\label{P(S,QE,QX)}
P_{\beta}(S, Q_{\varepsilon}, Q_x) \propto
\left( \det Q_{\varepsilon} \right)^{-2\beta N-3(1-\beta/2)} \nonumber \\
\times
\exp \left\{ -\frac{\beta}2 \; {\rm tr}
\left[ Q_{\varepsilon}^{-1} + \frac 18
\left( Q_{\varepsilon}^{-1} Q_x \right)^2 \right] \right\} .
\end{eqnarray}
$S$ is independent of $Q_{\varepsilon}$ and $Q_x$, and uniformly distributed in the
space of scattering matrices. Following \cite{BrouwerPRL1997}, $Q_x$ has a Gaussian
distribution with a width set by $Q_{\varepsilon}$, that can be parametrized as
follows \cite{BrouwerPRL1997}
\begin{equation}\label{paraQX}
Q_x = {\Psi^{-1}}^{\dagger} H \Psi^{-1},
\end{equation}
where $\Psi$ is a $N\times N$ matrix, complex in the unitary case and real in the
orthogonal one, such that
\begin{equation}\label{paraQE}
Q_{\varepsilon} = {\Psi^{-1}}^{\dagger} \Psi^{-1},
\end{equation}
and $H$ is a $N\times N$ Hermitian matrix for $\beta=2$, and real symmetric for
$\beta=1$. $H$ has a Gaussian distribution with zero mean and a variance
\begin{equation}\label{varH}
\left\langle H_{ab} H_{cd} \right\rangle = \left\{ \begin{array}{ll}
4 \, \delta_{ad} \delta_{bc} & \beta=2 \\
4 \left( \delta_{ad} \delta_{bc} + \delta_{ac} \delta_{bd} \right) &
\beta=1 \end{array} \right. \, ,
\end{equation}
as can be seen by substituting (\ref{paraQX}) and (\ref{paraQE}) into
(\ref{P(S,QE,QX)}). Now, we diagonalize $Q_{\varepsilon}$,
\begin{equation}\label{Qdiag}
Q_{\varepsilon} = W {\hat \tau} W^{\dagger} .
\end{equation}
The elements $\{\tau_n\}$ ($n=1,\ldots,N$) of ${\hat\tau}$ are the dimensionless
delay times. Their reciprocals $x_n=1/\tau_n$ ($n=1,\ldots,N$) are distributed
according to the Laguerre ensemble \cite{BrouwerPRL1997},
\begin{equation}\label{Laguerre}
P_L^{(\beta)} \left(x_1,\ldots,x_N \right) \propto
\prod_{a<b} \left| x_a - x_b \right|^{\beta}
\prod_c x_c^{\beta N/2} e^{-\beta x_c/2} \, .
\end{equation}
The matrix of eigenvectors, $W$, is uniformly distributed in the unitary
(orthogonal) group for $\beta=2$ ($\beta=1$).

For the calculations we are interested here, it is also convenient to parametrize
the $S$ matrix and its parametric derivative as \cite{Cremers}
\begin{equation}\label{paraSdqS}
S = UV , \quad
\frac{\partial S}{\partial\varepsilon} = i\, U Q_{\varepsilon} V , \quad
\frac{\partial S}{\partial x} = i\, U Q_x V ,
\end{equation}
where $U$, $V$ are the most general $N\times N$ unitary matrices in the
unitary case ($\beta=2$), while $V=U^T$ in the orthogonal one ($\beta=1$).


\subsection{Chaotic scattering by symmetric cavities in the presence
of absorption}
\label{subsec:theory-symm}

\begin{figure}
\includegraphics[width=6.0cm]{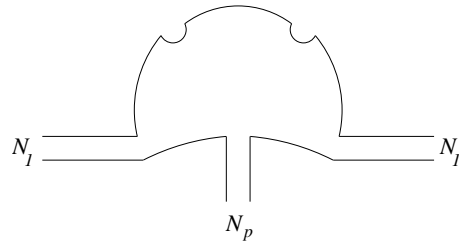}
\caption{A ballistic chaotic symmetric cavity connected to two leads
supporting $N_1$ channels. $N_p$ non transmitting channels are attached
to the cavity to model the absorption.}
\label{fig:scavity}
\end{figure}

For a system with spatial left-right (LR) symmetry, as shown in Fig.
\ref{fig:scavity}, the $S$ matrix is block diagonal in a basis of definite parity
with respect to reflections, with a circular ensemble in each block
\cite{Gopar1996,Baranger1996}.

In the presence of absorption the $S$ matrix that describes the scattering of
LR ballistic cavity connected to two waveguides, is of dimension $N=2N_1+N_p$,
where $N_1$ are the number of channels in each waveguide (the two waveguides have
the same number of channels and are symmetrically positioned); $N_p$ is the
number of absorption channels that we assume symmetrically distributed in the
cavity. In this case, the general structure for $S$ is \cite{Baranger1996}
\begin{equation} \label{Symmstruct}
S = \left(
\begin{array}{cc}
r' & t' \\ t' & r'
\end{array}
\right) ,
\end{equation}
where $r'$, $t'$ are $N'\times N'$ matrices, with $N'=N_1+N_p/2$. They represent
the reflection and transmission matrices, respectively, associated to the total
$S$ matrix given by (\ref{Symmstruct}), and not for the physical one. The
$N_1\times N_1$ transmission and reflection matrices, $t$ and $r$, associated to
the system with absorption, are submatrices of $t'$ and $r'$, respectively.

$S$-matrices of the form given by Eq. (\ref{Symmstruct}), which also satisfy
(\ref{unitary}) are appropriate for systems with reflection symmetry in the
absence of TRI. With the additional condition (\ref{unitary-symmetric}) it is
appropriate for LR-systems in the presence of TRI \cite{Mello2004}. However, when
TRI is broken by a uniform magnetic field, the problem of LR-symmetric cavities is
mapped \cite{Baranger1996} to the one of asymmetric cavities with $\beta=1$ with
$t'$ replaced by $r'$.

Matrices with the structure (\ref{Symmstruct}) can be brought to a
block-diagonal form \cite{Gopar1996}
\begin{equation} \label{block-diagonal}
S = R_0^T \left(
\begin{array}{cc}
S_1 & 0 \\ 0 & S_2
\end{array}
\right) R_0 \, ,
\end{equation}
where $R_0$ is the rotation matrix
\begin{equation} \label{rotation}
R_0 = \frac{1}{\sqrt{2}} \left(
\begin{array}{rr}
\openone_{N'} & \openone_{N'} \\ -\openone_{N'} & \openone_{N'}
\end{array}
\right) ,
\end{equation}
$\openone_n$ denotes the $n\times n$ unit matrix; $S_1=r'+t'$, $S_2=r'-t'$ are
the most general $N'\times N'$ scattering matrices. They are statistically
uncorrelated and uniformly distributed: CUE ($\beta=2$), COE ($\beta=1$)
\cite{Gopar1996}.

The transmission and reflection coefficients $T$ and $R$, for LR-symmetric
ballistic cavity in the presence of absorption are then given by
\begin{eqnarray}
T & = & \frac{1}{4} \sum_{a,b=1}^{N_1}
\left| (S_1)_{ab} - (S_2)_{ab} \right|^2 \quad \text{and}
\label{TSymm} \\
R & = & \frac{1}{4} \sum_{a,b=1}^{N_1}
\left| (S_1)_{ab} + (S_2)_{ab} \right|^2 \,,
\label{RSymm}
\end{eqnarray}
respectively.

The parametric derivative of $S$ is defined through the parametric derivatives
of $S_1$ and $S_2$ as in Eqs. (\ref{dES}) and (\ref{dXS}). The joint distribution
(\ref{P(S,QE,QX)}) is satisfied for each matrix $S_j$ ($j=1$, 2). Finally, we note
that they can be parametrized as in Eqs. (\ref{paraSdqS}).

In what follows we calculate the mean and variance of $\partial T/\partial q$
and $\partial R/\partial q$, where by $q$ we mean $\varepsilon$ or $x$. Also, we
calculate the correlations between the $q$-derivative of the channel-channel
transmission coefficients.


\section{Mean and variance of $\partial T/\partial q$ and
$\partial R/\partial q$ ($q=\varepsilon,\, x$) for asymmetric cavities}
\label{sec:asymm-fluc}

In this section we first calculate the mean of the $q$-derivative
($q=\varepsilon,\,x$) of $T$ and $R$. Second, we calculate correlation coefficient
between the $q$-derivative of two channel-channel transmission coefficients, from
where, finally, we can obtain the variance of $\partial T/\partial q$ and
$\partial R/\partial q$. The present section is devoted to asymmetric cavities for
both $\beta=1$ and $\beta=2$ symmetries.

By convenience we define the probability to go from channel $b$ to channel $a$ as
\begin{equation} \label{probab}
\sigma_{ab} = \left| S_{ab} \right|^2 \, ;
\end{equation}
from Eqs. (\ref{TRAs}) we can write
\begin{eqnarray}
\frac{\partial T}{\partial q} & = &
\sum_{a\in 1} \sum_{b\in 2}
\frac{\partial \sigma_{ab}}{\partial q} \quad \text{and} \label{dqT-asymm} \\
\frac{\partial R}{\partial q} & = & \sum_{a,b\in 1}
\frac{\partial \sigma_{ab}}{\partial q} \label{dqR-asymm} \, .
\end{eqnarray}

The ensemble average of $\partial T/\partial q$ and $\partial R/\partial q$ can be
calculated if we substitute the parametrization (\ref{paraSdqS}) into Eqs.
(\ref{dqT-asymm}) and (\ref{dqR-asymm}). In this way, we get expressions in terms
of twice the real part of products of averages of linear expressions in $Q_q$ times
averages of nonlinear expressions in $V$ and/or $U$ ($V=U^T$ for $\beta=1$).
Using the results of Ref. \onlinecite{Mello1990}, the averages with respect to
$U$ or $V$ are real positive numbers, while $\langle(Q_x)_{ab}\rangle=0$
because the matrix $H$ of Eq. (\ref{paraQX}) has zero mean;
$\langle(Q_{\varepsilon})_{ab}\rangle$ is a purely imaginary. Then, the
results are
\begin{equation}
\langle\partial T/\partial q\rangle = 0 = \langle\partial R/\partial q\rangle,
\quad q=\varepsilon,\, x,
\end{equation}
as expected because the distributions of $\partial T/\partial q$ and
$\partial R/\partial q$ are symmetric with respect to the zero derivative
\cite{Brouwer1997,mc}.

The fluctuations require a more sophisticated analysis. Let us define the
correlation coefficients by
\begin{equation} \label{Cqab-1}
{C_q^{(\beta)}}_{a'b'}^{ab} =
\left\langle \frac{\partial \sigma_{ab}}{\partial q}
\frac{\partial \sigma_{a'b'}}{\partial q}
\right\rangle  .
\end{equation}
The variances of $\partial T/\partial q$ and $\partial R/\partial q$ are then given
by
\begin{eqnarray}
\left\langle \left( \partial T/\partial q \right)^2\right\rangle
& = & \sum_{a,a'\in 1} \sum_{b,b'\in 2} {C_q^{(\beta)}}_{a'b'}^{ab}
\label{dqT2} \\
\left\langle \left( \partial R/\partial q \right)^2\right\rangle
& = & \sum_{a,a'\in 1} \sum_{b,b'\in 1} {C_q^{(\beta)}}_{a'b'}^{ab}
\label{dqR2} \, ,
\end{eqnarray}
with
\begin{eqnarray}\label{Cqab-2}
{C_q^{(\beta)}}_{a'b'}^{ab} & = & 2 \text{Re} \left[
\left\langle S_{ab} S_{a'b'}^* \frac{\partial S_{ab}^*}{\partial q}
\frac{\partial S_{a'b'}}{\partial q} \right\rangle
\right. \nonumber \\ & + & \left.
\left\langle S_{ab}^* S_{a'b'}^* \frac{\partial S_{ab}}{\partial q}
\frac{\partial S_{a'b'}}{\partial q} \right\rangle \right] ,
\end{eqnarray}
where we have written explicitly the elements of the $S$ matrix.

Because of the complexity of the calculations, in the rest of this section we
will consider the two symmetries $\beta=1$ and $\beta=2$ in a separate way.


\subsection{The orthogonal case}
\label{subsec:asymm-beta1}


\subsubsection{The correlator ${C_q^{(1)}}_{a'b'}^{ab}$}

In the orthogonal case, the substitution of the parametrization given by Eq.
(\ref{paraSdqS}), with $V=U^T$, into Eq. (\ref{Cqab-2}) gives the result
\begin{eqnarray} \label{Cqab(1)}
{C_q^{(1)}}_{a'b'}^{ab} & = & 2 \, \text{Re}
\sum_{\alpha,\beta=1}^N \sum_{\alpha',\beta'=1}^N
\left\langle (Q_q)_{\alpha\beta} (Q_q)_{\alpha'\beta'} \right\rangle
\nonumber \\ & \times &
\sum_{c,c'=1}^N \left[
J \left( \alpha,\beta,c,c \right) - J \left( c,c,\alpha,\beta \right)
\right] , \quad
\end{eqnarray}
where, in order to simplify the expression, we have defined the coefficients
\begin{eqnarray} \label{MUU*}
& & J(\alpha,\beta,\gamma, \delta) \equiv
M_{a\gamma,b\delta,a'\alpha',b'\beta'}^{a\alpha,b\beta,a'c',b'c'}
\nonumber \\ & &
\equiv \left\langle
U_{a\gamma} U_{b\delta} U_{a'\alpha'} U_{b'\beta'}
U_{a\alpha}^* U_{b\beta}^* U_{a'c'}^* U_{b'c'}^*
\right\rangle .
\end{eqnarray}
The first (last) two places $\alpha$, $\beta$ ($\gamma$, $\delta$) of the argument
of $J(\alpha, \beta, \gamma, \delta)$, refers to the second and fourth positions in
the upper (lower) indices of
$M_{a\gamma,b\delta,a'\alpha',b'\beta'}^{a\alpha,b\beta,a'c',b'c'}$ which is
defined by the second line of Eq. (\ref{MUU*}). As we can see in App. \ref{app:Ms},
the rest of the indices of the coefficients $M$ are not modified in the
construction of Eq. (\ref{Cqab(1)}). Those coefficients $M$ were calculated in Ref.
\onlinecite{Mello1990} [see Eq. (6.3) of that reference]; we apply those results to
our particular case in App. \ref{app:Ms}. The sums with respect $c$, $c'$ appearing
in the second line of Eq. (\ref{Cqab(1)}) give the result
\begin{eqnarray} \label{M2-M1}
& & \sum_{c,c'=1}^N \left[
J(\alpha,\beta,c,c) - J(c,c,\alpha,\beta)
\right] \nonumber \\ & & =
- n_1 \, \delta_{\alpha}^{\beta} \delta_{\alpha'}^{\beta'}
- n_2 \, \delta_{\alpha}^{\alpha'} \delta_{\beta}^{\beta'}
+ n_3 \, \delta_{\alpha}^{\beta'} \delta_{\beta}^{\alpha'} \, ,
\end{eqnarray}
We substitute Eq. (\ref{M2-M1}) into Eq. (\ref{Cqab(1)}) and simplifly
to obtain a result that depends on $n_1$ and $n_3-n_2$. In App. \ref{app:Ms}
we show that $n_3-n_2=Nn_1$ [see Eq. (\ref{n3-n2})] where $n_1$ is given
by Eq. (\ref{n1-3}). Then, we write Eq. (\ref{Cqab(1)}) as
\begin{equation} \label{Cqab(1)-2}
{C_q^{(1)}}_{a'b'}^{ab} = 2\, n_1 \, \text{Re} \, K_q^{(1)} \, ,
\end{equation}
where
\begin{equation} \label{Kq-1}
K_q^{(1)} = N \sum_{\alpha=1}^N
\left\langle \left( Q_q^2 \right)_{\alpha\alpha} \right\rangle -
\sum_{\alpha,\beta=1}^N \left\langle (Q_q)_{\alpha\alpha}
(Q_q)_{\beta\beta} \right\rangle .
\end{equation}

$K_{\varepsilon}^{(1)}$ is given by Eq. (\ref{Kq-1}) with $q$ replaced by
$\varepsilon$. $K_x^{(1)}$ can be written in terms of $Q_{\varepsilon}$ by direct
substitution of Eq. (\ref{paraQX}) into Eq. (\ref{Kq-1}). The average over the
matrix $H$ is performed taking into account Eqs. (\ref{varH}) and (\ref{paraQE})
for $\beta=1$; the result is
\begin{equation}
K_x^{(1)} = 4 ( N - 2 ) \sum_{\alpha=1}^N
\left\langle (Q_{\varepsilon}^2)_{\alpha\alpha} \right\rangle +
4N \sum_{\alpha,\beta=1}^N
\left\langle (Q_{\varepsilon})_{\alpha\alpha}
(Q_{\varepsilon})_{\beta\beta} \right\rangle .
\end{equation}
Now, we use the diagonal form of $Q_{\varepsilon}$, Eq. (\ref{Qdiag}). $K_q^{(1)}$
becomes independent of the unitary matrix $W$, and depends on two eigenvalues of
$Q_{\varepsilon}$ as
\begin{eqnarray}
K_x^{(1)} & = & 4 N ( N - 1 ) \left[
2 \left\langle \tau_1^2 \right\rangle +
N \left\langle \tau_1 \tau_2 \right\rangle \right] \label{Kxbeta1-2}, \\
K_{\varepsilon}^{(1)} & = & N ( N - 1 ) \left[
\left\langle \tau_1^2 \right\rangle -
\left\langle \tau_1 \tau_2 \right\rangle \right] \label{Kebeta1-2}.
\end{eqnarray}
The remaining averages of the $\tau$ variables are performed by direct
integration using Eq. (\ref{Laguerre}) for $\beta=1$. $\langle\tau_1^2\rangle$
diverges for $N=1$, while the next four values of $N$ give the general term
\begin{equation}\label{tau-av-beta1}
\left\langle\tau_1^2\right\rangle = \frac{2 N!}
{ \left( N-2 \right) \left( N+1 \right) !}\, , \, \quad
\langle \tau_1 \tau_2 \rangle = \frac{ \left( N-1 \right)!}
{\left( N+1 \right) !} .
\end{equation}
Then, Eqs. (\ref{Kxbeta1-2}) and (\ref{Kebeta1-2}) are written as
\begin{equation} \label{KXKEb1}
K_x^{(1)} = 4 N K_{\varepsilon}^{(1)} \, , \qquad
K_{\varepsilon}^{(1)} = \frac{(N-1)(N+2)}{(N-2)(N+1)} .
\end{equation}

Eqs. (\ref{n1-3}), (\ref{Cqab(1)-2}), (\ref{KXKEb1}) are combined to give the
desired results for the correlation coefficients, namely
\begin{eqnarray}
{C_x^{(1)}}_{a'b'}^{ab} & = & 4 N {C_{\varepsilon}^{(1)}}_{a'b'}^{ab}
\label{CX(1)} \\
{C_{\varepsilon}^{(1)}}_{a'b'}^{ab} & = &
\frac 2{ ( N-2 ) N^2 ( N+1 )^2 ( N+3 ) }
\nonumber \\ & \times &
\left\{ 2 ( 1 + \delta_a^b )
( 1 + \delta_{a'}^{b'} ) \right.
\nonumber \\ & + &
( N+1 ) ( N+2 )
( \delta_a^{a'} \delta_b^{b'} +
\delta_a^{b'} \delta_b^{a'} )^2
\nonumber \\ & - &
( N+1 ) \left[
\delta_b^{a'} + \delta_b^{b'} + \delta_a^{a'} +\delta_a^{b'} \right.
\label{CE(1)} \\ & & +
2 \, \delta_a^b \delta_{a'}^{b'}
( \delta_b^{b'} \delta_{a'}^a +
\delta_b^{a'} \delta_{b'}^a ) +
2 ( \delta_b^{b'} \delta_{a'}^b \delta_{b'}^{a'}
\nonumber \\ & & + \left. \left.
\delta_a^b \delta_b^{a'} \delta_{a'}^a +
\delta_a^b \delta_b^{b'} \delta_{b'}^a +
\delta_a^{b'} \delta_{a'}^a \delta_{b'}^{a'} )
\right] \right\} \nonumber ,
\end{eqnarray}
where the dependence on the absorption strength $\gamma_p=N_p$ is through
$N=N_1+N_2+N_p$.

From Eqs. (\ref{CX(1)}) and (\ref{CE(1)}) we analyze several cases of
interest. First, $a'=a\in 1$, $b'=b\in 2$, give the variances (maximal
correlations) of the energy and parametric derivatives of the channel-channel
transmission coefficient $\partial\sigma_{ab}/\partial q$
($q = \varepsilon,\, x$); those are
\begin{eqnarray}
\left\langle\left( \partial\sigma_{ab}/\partial x \right)^2\right\rangle
& = & 4 N
\left\langle\left(\partial\sigma_{ab}/\partial\varepsilon\right)^2\right\rangle
 \label{vardXTab(1)} \\
\left\langle\left(\partial\sigma_{ab}/\partial\varepsilon\right)^2\right\rangle
& = & \frac{ 2( N^2 + N + 2) }
{ (N-2) N^2 (N+1)^2 (N+3)} \label{vardETab(1)} .
\end{eqnarray}
We see that for strong absorption, $\gamma_p=N_p\gg N1,\, N2$, they behave as
\begin{equation} \label{vardqTab(1)-2}
\left\langle\left(
\partial\sigma_{ab}/\partial x
\right)^2\right\rangle \sim \gamma_p^{-3}, \qquad
\left\langle\left(
\partial\sigma_{ab}/\partial\varepsilon
\right)^2\right\rangle \sim \gamma_p^{-4} .
\end{equation}
Second, when $a'=a\in 1$ and $b,\,b'\in 2$, but $b'\neq b$, in the limit of strong
absorption we obtain
\begin{equation} \label{uncor-dqTab(1)-1}
\left\langle
\frac{\partial\sigma_{ab}}{\partial x}
\frac{\partial\sigma_{ab'}}{\partial x}
\right\rangle \sim \gamma_p^{-4}, \qquad
\left\langle
\frac{\partial\sigma_{ab}}{\partial\varepsilon}
\frac{\partial\sigma_{ab'}}{\partial\varepsilon}
\right\rangle \sim \gamma_p^{-5},
\end{equation}
that are smaller compared with the variances given by Eqs. (\ref{vardqTab(1)-2})
by a factor of $\gamma_p^{-1}$. Finally, when all the indices are different, in
the limit of strong absorption, the correlator between the parametric derivatives
of two different single channel transmission coefficients behaves as
\begin{equation} \label{uncor-dqTab(1)-2}
\left\langle
\frac{\partial\sigma_{ab}}{\partial x}
\frac{\partial\sigma_{a'b'}}{\partial x}
\right\rangle \sim \gamma_p^{-5}, \qquad
\left\langle
\frac{\partial\sigma_{ab}}{\partial\varepsilon}
\frac{\partial\sigma_{a'b'}}{\partial\varepsilon}
\right\rangle \sim \gamma_p^{-6} ,
\end{equation}
which are $\gamma_p^{-2}$ times the variances.
We conclude that for strong absorption, up to the order of
$\langle(\partial\sigma_{ab}/\partial q)^2\rangle$, the correlations
between the elements $\partial\sigma_{ab}/\partial q$, for $a\in 1$, $b\in 2$,
are very small. Those quantities enter in the construction of
$\partial T/\partial q$ [see Eq. (\ref{dqT-asymm})] and can be treated as
$N_1N_2$ uncorrelated variables with the same distribution. This is a relevant
simplification when the distribution of the parametric derivative of the total
transmission coefficient is desired, assuming the one for each
$\partial\sigma_{ab}/\partial q$ is known. That is the case of Ref.
\onlinecite{Schanze2003} where the numerical evidence shows an exponential
decay for $P_1(\partial\sigma_{ab}/\partial\varepsilon)$,
$P_1(\partial T/\partial\varepsilon)$ being calculated in a very straightforward
manner. Eqs. (\ref{vardXTab(1)}) and (\ref{vardETab(1)}) can be used to obtain the
decay constant as a function of $\gamma_p$ \cite{Schanze2003}.


\subsubsection{Statistical fluctuations of $\partial T/\partial q$ and
$\partial R/\partial q$ ($q=\varepsilon, \, x$)}

The second moment of the distribution of $\partial T/\partial q$ is obtained
from Eqs. (\ref{CX(1)}) and (\ref{CE(1)}) by direct subtitution into Eq.
(\ref{dqT2}); we obtian
\begin{eqnarray}
\left\langle\left( \partial T/\partial x \right)^2\right\rangle & = &
4 N \left\langle\left( \partial T/\partial\varepsilon \right)^2\right\rangle
\label{vardXT(1)} \\
\left\langle\left( \partial T/\partial\varepsilon \right)^2\right\rangle & = &
2 \frac{ N_1 N_2 \left[ (N+1)(\gamma_p+2) + 2 N_1 N_2 \right] }
{(N-2) N^2 (N+1)^2 (N+3)} \label{vardET(1)} \nonumber \\
\end{eqnarray}

For the particular case $N_1=N_2=1$, Eqs. (\ref{vardXT(1)}) and (\ref{vardET(1)})
reduce to Eqs. (\ref{vardXTab(1)}) and (\ref{vardETab(1)}), respectively. Also,
when $\gamma_p=N_p=0$, which means no absorption, the variance of
$\partial T/\partial q$ diverges. This is in agreement with Ref.
\onlinecite{Brouwer1997} where the distribution of $\partial T/\partial q$, was
obtained in the absence of absorption. The distribution has long tails and a
divergent second moment. This divergence is suppressed in the presence of
absorption. We also see that the divergence of
$\langle(\partial T/\partial q)^2\rangle$ disappear when $N_1$ or $N_2$ is larger
than one for any absorption strength.

In similar way, we substitute Eqs. (\ref{CX(1)}) and (\ref{CE(1)}) into Eq.
(\ref{dqR2}) to obtain
\begin{eqnarray}
\left\langle\left( \partial R/\partial x \right)^2\right\rangle & = &
4 N \left\langle\left( \partial R/\partial\varepsilon \right)^2\right\rangle
\label{dXRAs-1} \\
\left\langle\left( \partial R/\partial\varepsilon \right)^2\right\rangle & = &
4 \frac{ N_1 (N_1+1) (N-N_1) (N-N_1+1) }
{ (N - 2) N^2 (N + 1)^2 (N + 3) }, \nonumber \\ \label{dERAs-1}
\end{eqnarray}
for $N\ge 2$, while $\langle(\partial R/\partial q)^2\rangle$ is infinite for
$N=1$ as $N\neq N_1$ or $N\neq N_1-1$. When $N=N_1$ or $N=N_1-1$,
$\langle(\partial R/\partial q)^2\rangle=0$.

Consider the case $N_1=1$ and $N_2=0$ ($N=1+N_p$), which is relevant to the
experimental data of Ref. \onlinecite{Mendez-Sanchez2003}. In the absence of
absorption $\gamma_p=N_p=0$, $N=N_1=1$ and
$\langle(\partial R/\partial q)^2\rangle=0$ as expected ($R=1$). For
$\gamma_p=N_p\le 1$, $\langle(\partial R/\partial q)^2\rangle$ is infinite.
Again, the divergence is suppresed for $\gamma_p=N_p>1$.

We recall that our results are valid in the strong absorption regime where
by convenience we assumed perfect coupling ($T_p=1$) of the $N_p$ absorbing
channels. The absorption strength $\gamma_p=N_p$ takes only integer values.
However, a simple extrapolation to $T_p<1$, which means arbitrary
$\gamma_p=T_pN_p$, works qualitatively well. In Fig. \ref{fig:agr12}a) we
compare Eq. (\ref{vardET(1)}), for $N_1=N_2=2$, with the results from numerical
simulations \cite{num} for $T_p=0.025$, 0.05, 0.075, 0.1, 0.125, and 0.15 with
$N_p=200$ which give $\gamma_p=5$, 10, 15, 20, 25, 30.

In the absence of absorption, i.e. $N_p=0$, the results presented here are strictly valid. In this case $R+T=1$ and the distribution of $\partial R/\partial q$ is equal to that of $\partial T/\partial q$. In particular their variances are the same: it is easy to verify that Eqs. (\ref{dXRAs-1}) and (\ref{dERAs-1}) reduce to Eqs. (\ref{vardXT(1)}) and (\ref{vardET(1)}), in complete agreement with the results obtained directly from the known distribution of those quantities in the absence of absorption \cite{Brouwer1997}. The particular cases $N_1=1$, $N_2=0$, and $N_1=N_2=1$ has been explained above. Similar conclusions are valid for the unitary case and for $\beta=1,2$ for reflection symmetric case below. 


\subsection{The unitary case}
\label{subsec:asymm-beta2}


\subsubsection{The correlator ${C_q^{(2)}}_{a'b'}^{ab}$}

The unitary case is simpler than the orthogonal one. Following the same
procedure, we substitute the parametrization (\ref{paraSdqS}) into Eq.
(\ref{Cqab-2}) with the result
\begin{eqnarray*}
{C_q^{(2)}}_{a'b'}^{ab} = 2 \, \text{Re}
\sum_{\alpha,\beta=1}^N \sum_{\alpha',\beta'=1}^N \sum_{c,c'=1}^N
\Bigg[ \left\langle (Q_q)_{\beta\alpha} (Q_q)_{\alpha'\beta'} \right\rangle
\nonumber \\ \times
M_{ac,a'\alpha'}^{a\alpha, a'c'} M_{cb,\beta' b'}^{\beta b,c'b'} -
\left\langle (Q_q)_{\alpha\beta} (Q_q)_{\alpha'\beta'} \right\rangle
M_{a\alpha,a'\alpha'}^{ac, a'c'} M_{\beta b,\beta' b'}^{cb,c'b'}
\Bigg],
\end{eqnarray*}
where we have defined
\begin{eqnarray} \label{Mabcd}
M_{ab,cd}^{a'b',c'd'} \equiv
\left\langle U'_{ab} U'_{cd} {U'}_{a'b'}^* {U'}_{c'd'}^* \right\rangle ,
\end{eqnarray}
with $U'$ a unitary matrix that denotes the unitary matrices $U$ or $V$ of Eq.
(\ref{paraSdqS}). Those coefficients have been
calculated in Ref. \onlinecite{Mello1990} and read
\begin{eqnarray} \label{Mabcd-2}
M_{ab,cd}^{a'b',c'd'} & = & \frac{1}{N^2-1} \left[ (
\delta_a^{a'}\delta_c^{c'} \delta_b^{b'}\delta_d^{d'} +
\delta_a^{c'}\delta_c^{a'} \delta_b^{d'}\delta_d^{b'} ) \right.
\nonumber \\ & - & \frac{1}{N} \left. (
\delta_a^{a'}\delta_c^{c'} \delta_b^{d'}\delta_d^{b'} +
\delta_a^{c'}\delta_c^{a'} \delta_b^{b'}\delta_d^{d'} )
\right] .
\end{eqnarray}

We substitute Eq. (\ref{Mabcd-2}) into ${C_q^{(2)}}_{a'b'}^{ab}$ and perform the
sum over the dummy indices, the result is
\begin{equation} \label{Cqab(2)-2}
{C_q^{(2)}}_{a'b'}^{ab} =
\frac{2 \left[ 1 - N ( \delta_a^{a'} + \delta_b^{b'} ) +
N^2\delta_a^{a'} \delta_b^{b'} \right]}{N^2(N^2-1)^2} \text{Re} \,
K_q^{(2)} \, ,
\end{equation}
where $K_q^{(2)}$ has the same form as Eq. (\ref{Kq-1}) but with the upper index 1
on the left-hand side replaced by 2, and the matrix $Q_q$ is an Hermitian one.
Again, $K_{\varepsilon}^{(2)}$ is obtained by replacing $q=\varepsilon$.
To write $K_x^{(2)}$ in terms of $Q_{\varepsilon}$ we use Eq. (\ref{paraQX}) and
perform the average over $H$ using Eq. (\ref{varH}) for $\beta=2$. The result is
\begin{equation}
K_x^{(2)} = -4 \sum_{\alpha=1}^N
\left\langle (Q_{\varepsilon}^2)_{\alpha\alpha} \right\rangle
+ 4 N \sum_{\alpha,\beta=1}^N
\left\langle (Q_{\varepsilon})_{\alpha\alpha}
(Q_{\varepsilon})_{\beta\beta} \right\rangle .
\end{equation}
Now, we substitute Eq. (\ref{Qdiag}) and perform the average over $W$ to obtain
\begin{eqnarray}
K_x^{(2)} & = & 4 N(N-1) \left[
\left\langle\tau_1^2\right\rangle +
N \left\langle \tau_1 \tau_2 \right\rangle \right] , \label{KX(2)} \\
K_{\varepsilon}^{(2)} & = & N(N-1) \left[
\left\langle\tau_1^2\right\rangle -
\left\langle \tau_1 \tau_2 \right\rangle \right] \label{KE(2)}
\end{eqnarray}
By direct integration of the first $N$ terms, Eq. (\ref{Laguerre}) for $\beta=2$
give
\begin{equation}\label{tau-averages-2}
\langle \tau_1^2 \rangle = \frac{2 N(N-2)!}{(N+1)!}\, , \, \quad
\langle \tau_1 \tau_2 \rangle = \frac{(N-1)!}{(N+1)!} .
\end{equation}
Eqs. (\ref{KX(2)}), (\ref{KE(2)}) and (\ref{tau-averages-2}) give
\begin{equation} \label{KXKEb2}
K_x^{(2)} = 4 N K_{\varepsilon}^{(2)} \, , \qquad
K_{\varepsilon}^{(2)} = 1 .
\end{equation}
Finally, we combine Eqs. (\ref{Cqab(2)-2}) and (\ref{KXKEb2}) with the result
\begin{eqnarray}
{C_x^{(2)}}_{a'b'}^{ab} & = & 4 N {C_{\varepsilon}^{(2)}}_{a'b'}^{ab}
\label{CXab(2)} \\
{C_{\varepsilon}^{(2)}}_{a'b'}^{ab} & = &
\frac{2\left[ 1 - N ( \delta_a^{a'} + \delta_b^{b'} ) +
N^2\delta_a^{a'} \delta_b^{b'} \right]}{N^2(N^2-1)^2}
\label{CEab(2)} .
\end{eqnarray}

Two different particular cases are of interest. The first one, a correlated case,
is obtained for $a'=a\in 1$ and $b'=b\in 2$, for which one obtains that
\begin{eqnarray}
\left\langle\left( \partial\sigma_{ab}/\partial x \right)^2\right\rangle & = &
4 N\left\langle\left(\partial\sigma_{ab}/\partial\varepsilon\right)^2\right\rangle,
\label{vardXTab(2)} \\
\left\langle\left( \partial\sigma_{ab}/\partial\varepsilon\right)^2\right\rangle
& = &
\frac{2}{ N^2 ( N + 1 )^2}, \label{vardETab(2)}
\end{eqnarray}
which for strong absorption they have the behavior given by Eq.
(\ref{vardqTab(1)-2}). Second, uncorrelated cases are obtained when $a'=a\in 1$,
$b'\neq b$ ($b,b'\in 2$), and when all the indices are different, in the strong
absorption limit. For large $\gamma_p=N_p$ Eqs. (\ref{uncor-dqTab(1)-1}) and
(\ref{uncor-dqTab(1)-2}) are also satisfied for $\beta=2$. Those quantities are
very small compared to the order of
$\langle(\partial\sigma_{ab}/\partial q)^2\rangle$, meaning that in this limit
the quantities $\partial\sigma_{ab}/\partial q$ for $a\in 1$ and $b\in 2$, can be
treated as $N_1N_2$ uncorrelated variables with the same distribution
$P_2(\partial\sigma_{ab}/\partial q)$. Numerical evidence \cite{Schanze2003} also
shows an exponential decay of $P_2(\partial\sigma_{ab}/\partial\varepsilon)$ for
strong absorption; the decay constant depends on $\gamma_p$ and can be obtained from the
variance of $\partial\sigma_{ab}/\partial q$.


\subsubsection{Fluctuations of $\partial T/\partial q$ and $\partial R/\partial q$
($q=\varepsilon, \, x$)}

The statistical fluctuations of the energy and parametric derivative of the total
transmission coefficient is obtained by direct substitution of Eqs. (\ref{CXab(2)})
and (\ref{CEab(2)}) into Eq. (\ref{dqT2}) for  $\beta=2$. The results are
\begin{eqnarray}
\left\langle \left( \partial T/\partial x\right)^2 \right\rangle & = & 4N
\left\langle \left( \partial T/\partial\varepsilon\right)^2 \right\rangle
\label{varX} \\
\left\langle \left( \partial T/\partial\varepsilon\right)^2 \right\rangle & = &
\frac{2 N_1 N_2 \left( N N_p + N_1 N_2 \right)}
{ N^2 \left( N^2 - 1\right)^2} \label{varE}
\end{eqnarray}

When $N_1=N_2=1$ we reproduce Eqs. (\ref{vardXTab(2)}) and (\ref{vardETab(2)}). In
this case, $\langle(\partial T/\partial q)^2\rangle$ does not diverges for
$\gamma_p=N_p=0$, in contrast with the $\beta=1$ case. Also, this agree with Ref.
\onlinecite{Brouwer1997}.

\begin{figure}
\includegraphics[width=7.0cm]{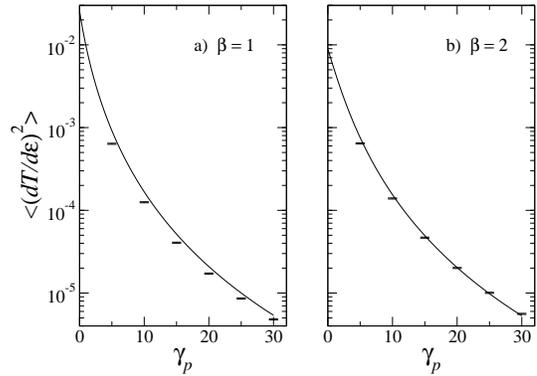}
\caption{$(\partial T/\partial\varepsilon)^2$ as a function of $\gamma_p=T_pN_P$
for an asymmetric cavity connected to two leads with $N_1=N_2=2$ open channels.
The errorbars indicates the result of numerical simulations \cite{num} for
$N_p=200$ and $T_p=0.025$, 0.05, 0.075, 0.1, 0.125, and 0.15, that give
$\gamma_p=5$, 10, 15, 20, 25, 30. The continuous line is the analytical formula
given by Eq. (\ref{vardET(1)}) for a) $\beta=1$, and Eq. (\ref{varE}) for b)  $\beta=2$.}
\label{fig:agr12}
\end{figure}

Although $\gamma_p$ takes only integer values, a simple extrapolation to non
integer values works qualitatively well as can be seen in Fig. \ref{fig:agr12} b),
where we have compared Eq. (\ref{varE}) for $N_1=N_2=2$, with the results from
numerical simulations \cite{num} for $T_p=0.025$, 0.05, 0.075, 0.1, 0.125, and 0.15
with $N_p=200$.

Similarly, we substitute Eqs. (\ref{CXab(2)}) and (\ref{CEab(2)}) into Eq.
Eq. (\ref{dqR2}) for $\beta=2$ to obtain the fluctuations of the derivative of
$R$; we have
\begin{eqnarray}
\left\langle \left(\partial R/\partial x\right)^2\right\rangle & = & 4N
\left\langle \left(\partial R/\partial\varepsilon\right)^2\right\rangle
\label{dXRAs(2)} \\
\left\langle \left(\partial R/\partial\varepsilon\right)^2\right\rangle & = &
\frac{2 N_1^2 ( N - N_1 )^2}{ N^2 ( N^2 - 1 )^2}.
\label{dERAs(2)}
\end{eqnarray}
In contrast with the $\beta=1$ case, $\langle(\partial R/\partial q)^2\rangle$
does diverges at all for $\beta=2$.


\section{Fluctuations of $\partial T/\partial q$ and $\partial R/\partial q$
($q=\varepsilon,\, x$) for symmetric cavities}
\label{sec:symm-fluc}

Because of the left-right symmetry of the cavity it is sufficient to consider
$\partial T/\partial q$, the results for $\partial R/\partial q$ are equivalent.
Also, as happens in asymmetric cavities, $\langle(\partial T/\partial x)^2\rangle$
is always $4N$ times $\langle(\partial T/\partial\varepsilon)^2\rangle$ [see Eqs.
(\ref{vardXT(1)}) and (\ref{varX})]. Then, we will concentrate on the variance of
the energy derivative of $T$.

For LR-symmetric cavities we define $\sigma'_{ab}$ as the channel-channel
transmission probability, i.e. the square modulus of each element $t'_{ab}$ of the
transmission matrix $t'$ of Eq. (\ref{Symmstruct}). It can be written as
\begin{equation} \label{sigmaSymm}
\sigma'_{ab} = \frac{1}{4} \left[ (\sigma_1)_{ab} + (\sigma_2)_{ab} -
2 \, \text{Re} \, f_{ab} \right] ,
\end{equation}
where the prime on the left hand side indicates that it is defined for LR-symmetric
cavities, while $\sigma_1$, $\sigma_2$ are defined by Eq. (\ref{probab}) and
correspond to $S_1$ and $S_2$ matrices; $f_{ab}$ is an interference term given by
\begin{equation}
f_{ab} = (S_1)_{ab} \, (S_2^*)_{ab} \, .
\end{equation}
The energy derivative of $T$ is given by
\begin{equation} \label{dqTSymm}
\partial T/\partial\varepsilon = \sum_{a,b=1}^{N_1}
\partial\sigma'_{ab}/\partial\varepsilon
\end{equation}
and its fluctuation by
\begin{equation} \label{vardqTSymm}
\left\langle\left(\partial T/\partial/\varepsilon\right)^2 \right\rangle =
\sum_{a,b=1}^{N_1} \sum_{a',b'=1}^{N_1}
{D^{(\beta)}_{\varepsilon}}_{a'b'}^{ab} \, ,
\end{equation}
where, analogous to Eq. (\ref{Cqab-1}) for $q=\varepsilon$, we have defined the
correlation coefficient for the symmetric case as
\begin{equation} \label{CqabSymm-0}
{D^{(\beta)}_{\varepsilon}}_{a'b'}^{ab} =
\left\langle
\frac{\partial\sigma'_{ab}}{\partial\varepsilon}
\frac{\partial\sigma'_{a'b'}}{\partial\varepsilon}
\right\rangle .
\end{equation}
Using Eq. (\ref{sigmaSymm}) we write Eq. (\ref{CqabSymm-0}) as
\begin{equation} \label{CqabSymm}
{D^{(\beta)}_{\varepsilon}}_{a'b'}^{ab} =
\frac{1}{8} \left[
{{C'}^{(\beta)}_{\varepsilon}}_{a'b'}^{ab} +
\text{Re} \, {F^{(\beta)}_{\varepsilon}}_{a'b'}^{ab} \right],
\end{equation}
where ${C'^{(\beta)}_{\varepsilon}}_{a'b'}^{ab}$ is given by Eq. (\ref{CE(1)}) for
$\beta=1$ and Eq. (\ref{CEab(2)}) for $\beta=2$, with $N$ replaced by
$N'=N_1+N_p/2$, while
\begin{equation}
{F^{(\beta)}_{\varepsilon}}_{a'b'}^{ab} = \left\langle
\frac{\partial f_{ab}}{\partial\varepsilon}
\frac{\partial f^*_{a'b'}}{\partial\varepsilon} \right\rangle .
\end{equation}

To arrive at Eq. (\ref{CqabSymm}) we used the fact that $S_1$ and $S_2$ are
statistically uncorrelated, equally and uniformly distributed such that
${C^{(\beta)}_{2\varepsilon}}_{a'b'}^{ab}={C^{(\beta)}_{1\varepsilon}}_{a'b'}^{ab}$,
that we define as ${{C'}^{(\beta)}_{\varepsilon}}_{a'b'}^{ab}$. Also, we use the
results $\langle[\partial(\sigma_1)_{ab}/\partial\varepsilon]
[\partial(\sigma_2)_{a'b'}/\partial\varepsilon]\rangle=0$ from one side,
$\langle[\partial(\sigma_j)_{ab}/\partial\varepsilon]
(\partial f_{a'b'}/\partial\varepsilon)\rangle=0$ ($j=1$, 2) for the other side, and
finally
$\langle(\partial f_{ab}/\partial\varepsilon)
(\partial f_{a'b'}/\partial\varepsilon)\rangle=0$ that are easily to verify.

In order to calculate ${F^{(\beta)}_{\varepsilon}}_{a'b'}^{ab}$ we write it
explicitly in terms of $S_1$, $S_2$; it is given by
\begin{eqnarray} \label{Fq}
{F^{(\beta)}_{\varepsilon}}_{a'b'}^{ab} = 2 \left[
\left\langle (S_1)_{ab} (S_1^*)_{a'b'} \right\rangle
\left\langle \frac{\partial (S_2^*)_{ab}}{\partial\varepsilon}
\frac{\partial (S_2)_{a'b'}}{\partial\varepsilon} \right\rangle \right.
\nonumber \\  +
\left.
\left\langle (S_1)_{ab}
\frac{\partial (S_1^*)_{a'b'}}{\partial\varepsilon}
\right\rangle
\left\langle (S_2)_{a'b'}
\frac{\partial (S_2^*)_{ab}}{\partial\varepsilon}
\right\rangle
\right].
\end{eqnarray}
The second line of the Eq. (\ref{Fq}) is zero as was shown in Ref.
\onlinecite{mc}.


\subsection{The $\beta=1$ symmetry}
\label{subsec:symm-beta1}


\subsubsection{The correlator ${D^{(1)}_{\varepsilon}}_{a'b'}^{ab}$}

From the appendix in Ref. \onlinecite{mc}, for $\beta=1$ we have
\begin{eqnarray}
\left\langle(S_1)_{ab}(S^*_1)_{a'b'}\right\rangle & = &
( \delta^{a'}_a\delta^{b'}_b + \delta^{b'}_a\delta^{a'}_b )/(N'+1) \label{SS*1} \\
\left\langle \frac{\partial (S_2^*)_{ab}}{\partial\varepsilon}
\frac{\partial (S_2)_{a'b'}}{\partial\varepsilon}\right\rangle & = &
\frac{ ( \delta_a^{a'} \delta_b^{b'} +
\delta_a^{b'} \delta_b^{a'} ) }{ N'(N'+1) }
\sum_{\alpha=1}^{N'}
\left\langle \left( Q_{\varepsilon}^2\right)_{\alpha\alpha} \right\rangle ,
\nonumber \\
\label{dqSdqS*-2}
\end{eqnarray}
where we have replaced $N$ by $N'=N_1+N_p/2$ and $X$ by $\varepsilon$. Then, after
we substitute Eqs. (\ref{SS*1}), (\ref{dqSdqS*-2}) and (\ref{Qdiag}) into Eq.
(\ref{Fq}) we perform the average over the unitary matrix $W$ to arrive to the
result
\begin{equation}
{F_{\varepsilon}^{(1)}}_{a'b'}^{ab} =
\frac{ 2 ( \delta_a^{a'} \delta_b^{b'} +
\delta_a^{b'} \delta_b^{a'} )^2 }{ (N'+1)^2 }
\left[\left\langle\tau_1^2\right\rangle -
\left\langle\tau_1\right\rangle^2\right].
\end{equation}
Eq. (\ref{Laguerre}) with $N$ replaced by $N'$ gives $\langle\tau_1\rangle=1/N'$
by direct integration, together with Eq. (\ref{tau-av-beta1}) lead us to the
result
\begin{equation}\label{FE}
{F_{\varepsilon}^{(1)}}_{a'b'}^{ab} =
\frac{ 2 ( N'^2 + N' + 2 )
( \delta_a^{a'} \delta_b^{b'} + \delta_a^{b'} \delta_b^{a'} )^2 }
{ ( N'-2 ) {N'}^2 ( N'+1 )^3 }.
\end{equation}
Finally, Eq. (\ref{CE(1)}) with $N'$ instead of $N$ and Eq. (\ref{FE}) gives the
result for ${D^{(1)}_{\varepsilon}}_{a'b'}^{ab}$ [see Eq. (\ref{CqabSymm})].

As for the asymmetric case, several cases are of particular interest.
A first correlated case is obtained when all indices are equal, which gives the
variance of the energy derivative of the transmission probability between
two channels symmetrically located, $\sigma'_{aa}$; it is
\begin{equation} \label{vardEsaa(1)}
\left\langle\left(
\partial\sigma'_{aa}/\partial\varepsilon
\right)^2\right\rangle =
\frac{N'({N'}^2 - 1) + (N'+3)({N'}^2+N'+2)}
{ (N'-2) {N'}^2 (N'+1)^3 (N'+3)}.
\end{equation}
A second correlated case is obtained for $a'=a$ and $b'=b$ but $a\neq b$, which
gives the energy derivative variance of the transmission coefficient
$\sigma'_{ab}$ between two channels not located in a symmetric way; we have
\begin{equation} \label{vardXsab(1)}
\left\langle\left(
\partial\sigma'_{ab}/\partial\varepsilon
\right)^2\right\rangle =
\frac{ [(N'+1)+(N'+3)]({N'}^2+N'+2) }
{ 4(N'-2){N'}^2(N'+1)^3(N'+3) }.
\end{equation}
The last two equations are different because of the reflection symmetry of the
cavity. At level of the matrices $S_1$ and $S_2$ [see Eq. (\ref{block-diagonal})],
the diagonal elements represent reflection amplitudes, while the off-diagonal ones
represent transmission amplitudes. In fact, the first term on the right hand side
of Eqs. (\ref{vardEsaa(1)}) and (\ref{vardXsab(1)}) are equal to Eqs.
(\ref{dERAs-1}) and (\ref{vardXTab(1)}) (except by a constant factor),
respectively, when $N_1=1$ and $N$ is replaced by $N'$. The second term of Eqs.
(\ref{vardEsaa(1)}) and (\ref{vardXsab(1)}) comes from interference between $S_1$
and $S_2$ [see Eq. (\ref{CqabSymm})].

In the limit of strong absorption,
$\langle(\partial\sigma'_{aa}/\partial\varepsilon)^2\rangle$ and
$\langle(\partial\sigma'_{ab}/\partial\varepsilon)^2\rangle$ behave as
$\gamma_p^{-4}$. In similar way, it is simple to verify that
$\langle(\partial\sigma'_{aa}/\partial\varepsilon)
(\partial\sigma'_{a'b'}/\partial\varepsilon)\rangle$ and
$\langle(\partial\sigma'_{ab}/\partial\varepsilon)
(\partial\sigma'_{ab'}/\partial\varepsilon)\rangle$ behave as $\gamma_p^{-5}$,
while $\langle(\partial\sigma'_{aa}/\partial\varepsilon)
(\partial\sigma'_{a'a'}/\partial\varepsilon)\rangle$,
$\langle(\partial\sigma'_{aa}/\partial\varepsilon)
(\partial\sigma'_{a'b'}/\partial\varepsilon)\rangle$, and
$\langle(\partial\sigma'_{ab}/\partial\varepsilon)
(\partial\sigma'_{a'b'}/\partial\varepsilon)\rangle$ go as $\gamma_p^{-6}$. As
happens in the asymmtric case, the variables $\partial\sigma'_{ab}/\partial q$, for
$a$, $b=1,\ldots,N_1$, are uncorrelated for strong absorption. They enter in the
construction of $\partial T/\partial q$ [see Eq. (\ref{dqTSymm})], the
distribution of which is easily obtained when the one for
$\partial\sigma'_{ab}/\partial q$ is known \cite{Schanze2003}.


\subsubsection{Variance of $\partial T/\partial\varepsilon$}

From Eqs. (\ref{CE(1)}) with $N$ replaced by $N'$, (\ref{FE}), (\ref{CqabSymm}) and
(\ref{vardqTSymm}) for $\beta=1$ we obtain the variance of the energy derivative of
$T$, the result is
\begin{eqnarray}
\left\langle\left(
\partial T/\partial\varepsilon
\right)^2\right\rangle & = &
\frac{ N_1 (N_1+1) }
{ 2 (N'-2) {N'}^2 (N'+1)^2 }
\Bigg[
\frac{ {N'}^2 + N' + 2 }{ N'+1 } \nonumber \\ & & +
\frac{ (N'-N_1) (N'-N_1+1) }{ N'+3 }
\Bigg]
\label{dETSymm}
\end{eqnarray}
The effect of the LR-symmetry is clear. The second term of the last
equation is similar to Eq. (\ref{dERAs-1}) with $R$ replaced by $T$. That is
because $\partial T/\partial q$ for LR-symmetric cavity has a similar expression
to $\partial R/\partial q$ for asymmetric cavity as can be seen by comparison of
Eq. (\ref{dqTSymm}) with Eq. (\ref{dqR-asymm}). The second term in Eqs.
(\ref{dETSymm}) comes from the interference term of matrices $S_1$, $S_2$
as explained above [see Eq. (\ref{CqabSymm})].

For $N_1=1$, $N'=1+N_p/2$, Eq. (\ref{dETSymm}) reduces to Eq. (\ref{vardEsaa(1)}).
In this case $\langle(\partial T/\partial q)^2\rangle$ diverges for
$\gamma_p=N_p\le2$, but remains finite for $\gamma_p=N_p>2$. When $N_1=2$
$\langle(\partial T/\partial q)^2\rangle$ diverges only for $\gamma_p=N_p=0$.
In both cases a complete agreement with the results of Ref. \onlinecite{mc} is
found.


\subsection{The $\beta=2$ symmetry}


\subsubsection{Correlations of $\partial\sigma_{ab}'/\partial q$}

Again, making an appropriate correspondence from Ref. \onlinecite{mc} we have
\begin{eqnarray}
\left\langle(S_1)_{ab}(S^*_1)_{a'b'}\right\rangle & = &
\delta^{a'}_a\delta^{b'}_b / N' \label{SS*2} \\
\left\langle \frac{\partial (S_2^*)_{ab}}{\partial\varepsilon}
\frac{\partial (S_2)_{a'b'}}{\partial\varepsilon}\right\rangle & = &
\frac{ \delta_a^{a'} \delta_b^{b'} }{ {N'}^2 }
\sum_{\alpha=1}^{N'}
\left\langle \left( Q_{\varepsilon}^2\right)_{\alpha\alpha} \right\rangle .
\label{dqSdqS*-2(2)}
\end{eqnarray}
We substitute Eqs. (\ref{SS*2}), (\ref{dqSdqS*-2(2)}) and (\ref{Qdiag}) into Eq.
(\ref{Fq}) for $\beta=2$, and perform the average over the unitary matrix $W$,
the result is
\begin{equation}\label{FX(2)}
{F^{(2)}_{\varepsilon}}_{a'b'}^{ab} =
\frac{ 2 ( {N'}^2+1 ) \delta_a^{a'} \delta_b^{b'} }
{ {N'}^4 ( {N'}^2 -1 ) },
\end{equation}
where we used Eq. (\ref{tau-averages-2}) and the result
$\langle\tau_1\rangle=1/N'$ which can be obtained by direct integration from
Eq. (\ref{Laguerre}). Finally, Eqs. (\ref{CEab(2)}) with $N$
replaced by $N'$, and (\ref{FX(2)}) gives the desired result for
${D^{(2)}_{\varepsilon}}^{ab}_{a'b'}$ [Eq. (\ref{CqabSymm}) for $\beta=2$].

In this $\beta=2$ symmetry there is not difference in the variance of the energy
derivative of channel-channel transmission coefficient whether the two single
channels are located symmetrically or not. It is given by
\begin{equation} \label{vardEsab(2)}
\left \langle \left(
\partial\sigma'_{ab}/\partial\varepsilon
\right)^2 \right \rangle =
\frac 1{4 {N'}^2 (N'+1)^2} +
\frac{ {N'}^2 +1 } { 4{N'}^4 ( {N'}^2-1 ) } .
\end{equation}
The first term on the right hand side is the same, except by a constant, as Eq.
(\ref{vardETab(2)}), replacing $N$ by $N'$. The second term comes from interference
between $S_1$ and $S_2$ [see Eq. (\ref{CqabSymm})]. For strong absorption,
$\langle(\partial\sigma'_{aa}/\partial\varepsilon)^2\rangle$ behaves as
$\gamma_p^{-3}$. Also, as $\gamma_p=N_p$ increases the quantities
$\partial\sigma'_{ab}/\partial\varepsilon$ for $a$, $b=1,\ldots,N_1$, become
uncorrelated.


\subsubsection{Variance of $\partial T/\partial\varepsilon$}

From Eqs. (\ref{CEab(2)}) with $N$ replaced by $N'$, (\ref{vardqTSymm}),
(\ref{CqabSymm}) for $\beta=2$, and (\ref{FX(2)}) we obtain
\begin{equation} \label{dETSymm(2)}
\left\langle\left(
\partial T/\partial\varepsilon
\right)^2 \right\rangle =
\frac{ N_1^2 ( N' - N_1 )^2 }
{ 4{N'}^2 ( {N'}^2 - 1 )^2 } +
\frac{ N_1^2 ( {N'}^2 + 1 ) }
{ 4{N'}^4 ( {N'}^2 - 1 ) }.
\end{equation}
Again, we note the effect of the LR-symmetry. The first term is similar to Eq.
(\ref{dERAs(2)}). The second term in Eq. (\ref{dETSymm(2)}) comes from the
interference term of matrices $S_1$, $S_2$.

For the one channel case ($N_1=1$),
$\langle(\partial T/\partial\varepsilon)^2\rangle$ diverges for $\gamma_p=N_p=0$,
in contrast with the asymmetric case for $\beta=2$, and in agreement with Ref.
\onlinecite{mc}. It remains finite for $\gamma_p>0$.

\begin{figure}
\includegraphics[width=7.0cm]{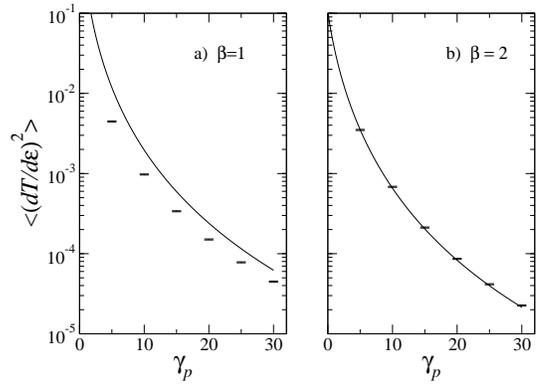}
\caption{The same as Fig. \ref{fig:agr12} but for symmetric cavities. The
continuous line is the analytical formula given by Eq. (\ref{dETSymm}) for a)  $\beta=1$, and Eq. (\ref{dETSymm(2)}) for b) $\beta=2$.}
\label{fig:sgr12}
\end{figure}

In Fig. \ref{fig:sgr12} we compare the analytical results (\ref{dETSymm}) and
(\ref{dETSymm(2)}), obtained with $T_p=1$, with the results from numerical
simulations for $T_p<1$; we observe a good qualitative agreement.


\subsection{TRI broken by a magnetic field}
\label{subsec:symm-Xbeta1X}

When TRI is broken by a magnetic field, the problem of a LR-symmetric cavity
is reduced to the problem of asymmetric cavity with $\beta=1$ symmetry but
the roles of $T$ and $R$ interchanged, such that the parametric derivative
of $T$ is given by Eq. (\ref{dqR-asymm}). All the elements
$\partial\sigma_{ab}/\partial q$, for $a$, $b=1,\ldots,N_1$, are uncorrelated
in the strong absorption limit.

In this case, for instance, the variance of $\partial T/\partial\varepsilon$ is
given by
\begin{equation}\label{XdETAs-2X}
\left\langle\left(
\partial T/\partial\varepsilon
\right)^2\right\rangle =
\frac{4 N_1 (N_1+1) (N-N_1) (N-N_1+1) }
{ (N - 2) N^2 (N + 1)^2 (N + 3) } .
\end{equation}

For a cavity connected to two leads each one supporting one open channel,
$\langle(\partial T/\partial\varepsilon)^2\rangle$ diverges for $\gamma_p=N_p=0$,
also in contrast with the $\beta=2$ case for asymmetric cavities.


\section{Summary and Conclusions}
\label{sec:conclusions}

The purpose of the present paper was the study of the statistical fluctuations of
the derivative of the transmission $T$ and reflection $R$ coefficients, with
respect to the incident energy $E$ and an external parameter $X$ (shape of the
cavity for instance), for ballistic chaotic cavities with absorption.

Our analytical results were obtained assuming $N_p$ equivalent absorbing channels
that are perfectly coupled to the cavity ($T_p=1$). This restrict our calculations
to be valid in the strong absorption limit, and the absorption strength takes
only integer values ($\gamma_p=N_p$). However, the results presented here are also
valid for no absorption, which means $\gamma_p=N_p=0$; they are in complete agreement with those obtained from known distributions of the parametric derivatives of $T$ and $R$ existing in the literature. Also, we have shown, by
comparison with numerical simulations, that a simple extrapolation to non integer
values of $\gamma_p$ is qualitatively correct.

We considered both asymmetric and left-right (LR) symmetric cavities connected to
two waveguides: $N_1$ channels on the left and $N_2$ channels on the right; both
symmetries, the presence and absence of time-reversal invariance
(TRI), were analyzed. For all cases, the fluctuations of the energy derivative are
smaller than those with respect to parametric. We found that
$\langle(\partial T/\partial x)^2\rangle=4M
\langle(\partial T/\partial\varepsilon)^2\rangle$, where
$\varepsilon=2\pi E/\Delta$ and $x=X/X_c$ with $\Delta$ the mean level spacing
and $X_c$ a typical scale for $X$. $M=N$ for asymmetric cavities, with
$N=N_1+N_2+N_p$, while $M=N/2$ for the symmetric case ($N_1=N_2$).

The correlation coefficient for the parametric derivative of the channel-channel
transmission probability $\sigma_{ab}$, $\partial\sigma_{ab}/\partial q$
($q=\varepsilon,\, x$), was calculated. It was shown that in the strong
absorption limit the different quantities $\partial\sigma_{ab}/\partial q$ for
become uncorrelated variables. They enter in the construction of
$\partial T/\partial q$. This is a relevant simplification when the distribution
$P(\partial T/\partial q)$ is desired assuming the one for
$\partial\sigma_{ab}/\partial q$ is known. That is the case of Ref.
\onlinecite{Schanze2003} where numerical simulations show evidence of an
exponential decay for $P(\partial\sigma_{ab}/\partial\varepsilon)$. The decay
constant $\lambda$ can be obtained directly from
$\langle(\partial T_{ab}/\partial\varepsilon)^2\rangle=2/\lambda^2$. A similar
behaviorfor $\partial\sigma_{ab}/\partial x$ is expected. This is in
contrast with the case of zero absorption where a long tail distribution is
obtained for the parametric conductance velocity \cite{Brouwer1997,mc}.

In the case of an asymmetric cavity connected to two leads each one with
one open channel ($N_1=N_2=1$), at zero absorption, we find that
$\langle(\partial T/\partial q)^2\rangle$ ($q=E$, $X$) is finite when no
TRI is present, but is infinite in the presence of TRI, in agreement with
Ref. \onlinecite{Brouwer1997} where a long tails distribution for
$\partial T/\partial q$ was obtained. The divergence in the second moment is
suppressed by absorption and we expect that the long tails become exponential
at sufficiently large $\gamma_p$ as mentioned above. This case also corresponds to
one of an asymmetric cavity with one-lead-one-channel ($N_1=1$, $N_2=0$) with one channel
of absorption perfectly coupled to the cavity, i.e $\gamma_p=1$. In this case,
$\langle(\partial R/\partial q)^2\rangle$ is infinite (finite) in the
presence (absence) of TRI. $\langle(\partial R/\partial q)^2\rangle=0$ at zero
absorption, as should be, and it is infinite for $0<\gamma_p<1$. The divergence
disappear for $\gamma_p>1$.

For a left-right (LR)-symmetric cavity connected to two waveguides with one open
channel each one ($N_1=N_2=1$), $\langle(\partial T/\partial q)^2\rangle$ is
divergent for $0\le\gamma_p\le 2$, and remains finite for $\gamma_p>2$ in the
presence of TRI. In the absence of TRI, the results are different in the presence
or absence of an applied magnetic field. However, in both cases
$\langle(\partial T/\partial q)^2\rangle$ diverges at $\gamma_p=0$, in contrast to
the asymmetric case, and in agreement with Ref. \onlinecite{mc}: a long tails
distribution for $\partial T/\partial q$ was found at zero absorption for presence
and absence of TRI.  $\langle(\partial T/\partial q)^2\rangle$ is finite for
$\gamma_p>0$. We also expect that the long tails will be suppressed at sufficiently
strong absorption \cite{Schanze2003}.

The results obtained in this paper help to understand some results presented in
Ref. \onlinecite{Schanze2003} about the energy derivative of the transmission
coefficient, and can serve as a motivation to extend that analysis to study the
distribution of the transmission derivative with respect to shape deformations, as
well as to motivate the analysis of the distribution of the parametric derivative
of the reflection coefficient.

\acknowledgments

The author thanks C. H. Lewenkopf for useful discussions and E. Casta\~no for
useful comments.


\appendix

\section{The coefficients $M(\alpha,\beta,\gamma,\delta)$}
\label{app:Ms}

Applying the result (6.3) of Ref. \onlinecite{Mello1990} to our case, we
can write Eq. (\ref{MUU*}) as
\begin{equation} \label{M(al,be)-1}
J(\alpha,\beta,\gamma,\delta) = A u_1 + B u_2 + C u_3 + D u_4 + E u_5 ,
\end{equation}
where 
\begin{eqnarray}\label{ABCDE} 
A & = & \frac{N^4-8N^2+6}{N^2(N^2-1)(N^2-4)(N^2-9)} \nonumber \\ 
B & = & -\frac{N(N^2-4)}{N^2(N^2-1)(N^2-4)(N^2-9)} \nonumber \\ 
C & = & \frac{2N^2-3}{N^2(N^2-1)(N^2-4)(N^2-9)}  \\ 
D & = & \frac{N^2+6}{N^2(N^2-1)(N^2-4)(N^2-9)} \nonumber \\ 
E & = & -\frac{5N}{N^2(N^2-1)(N^2-4)(N^2-9)} . \nonumber 
\end{eqnarray} 
and 
\begin{eqnarray} \label{u1} 
u_1 & = & 
a_1 ( \delta_{\gamma}^{\alpha} \delta_{\delta}^{\beta}
\delta_{\alpha'}^{c'} \delta_{\beta'}^{c'} ) + 
a_2 ( \delta_{\gamma}^{\alpha} \delta_{\delta}^{c'} 
\delta_{\alpha'}^{\beta} \delta_{\beta'}^{c'} )  
\nonumber \\ & + &
a_3 ( \delta_{\gamma}^{\alpha} \delta_{\delta}^{c'} 
\delta_{\alpha'}^{c'} \delta_{\beta'}^{\beta} ) + 
a_4 ( \delta_{\gamma}^{\beta} \delta_{\delta}^{\alpha} 
\delta_{\alpha'}^{c'} \delta_{\beta'}^{c'} ) 
\nonumber \\ & + & 
a_5 ( \delta_{\gamma}^{\beta} \delta_{\delta}^{c'} 
\delta_{\alpha'}^{\alpha} \delta_{\beta'}^{c'} ) + 
a_6 ( \delta_{\gamma}^{\beta} \delta_{\delta}^{c'}
\delta_{\alpha'}^{c'} \delta_{\beta'}^{\alpha} ) 
\nonumber \\ & + & 
a_7 ( \delta_{\gamma}^{c'} \delta_{\delta}^{\alpha} 
\delta_{\alpha'}^{\beta} \delta_{\beta'}^{c'} ) + 
a_8 ( \delta_{\gamma}^{c'} \delta_{\delta}^{\alpha} 
\delta_{\alpha'}^{c'} \delta_{\beta'}^{\beta} ) 
\\ & + & 
a_9 ( \delta_{\gamma}^{c'} \delta_{\delta}^{\beta}
\delta_{\alpha'}^{\alpha} \delta_{\beta'}^{c'} ) + 
a_{10} ( \delta_{\gamma}^{c'} \delta_{\delta}^{\beta}
\delta_{\alpha'}^{c'} \delta_{\beta'}^{\alpha} ) 
\nonumber \\ & + & 
a_{11} ( \delta_{\gamma}^{c'} \delta_{\delta}^{c'}
\delta_{\alpha'}^{\alpha} \delta_{\beta'}^{\beta} ) + 
a_{12} ( \delta_{\gamma}^{c'} \delta_{\delta}^{c'} 
\delta_{\alpha'}^{\beta} \delta_{\beta'}^{\alpha} ) 
\, , \nonumber
\end{eqnarray} 
with 
\begin{equation} \label{ajs} 
\begin{array}{ll} 
a_1 = 1 + \delta_{a'}^{b'} &  
a_2 = ( 1 + \delta_b^{b'} \delta_{b'}^{a'} ) \delta_b^{a'} \\
a_3 = ( 1 + \delta_b^{a'} \delta_{a'}^{b'}  ) \delta_{b'}^b &  
a_4 = ( 1 + \delta_{a'}^{b'} ) \delta_a^b \\ 
a_5 = ( \delta_b^{a'} + \delta_b^{b'} \delta_{b'}^{a'} ) 
\delta_a^b \delta_a^{a'} &  
a_6 = ( \delta_b^{b'} + \delta_b^{a'} \delta_{a'}^{b'} ) \delta_a^b 
\delta_{b'}^a \\ 
a_7 = ( \delta_a^{a'} + \delta_a^{b'} \delta_{b'}^{a'} ) \delta_b^a 
\delta_{a'}^b &
a_8 = ( \delta_a^{b'} + \delta_a^{a'} \delta_{a'}^{b'} ) \delta_b^a 
\delta_{b'}^b \\
a_9 = ( 1 + \delta_a^{b'} \delta_{b'}^{a'} ) \delta_a^{a'} &  
a_{10} = ( 1 + \delta_a^{a'} \delta_{a'}^{b'} ) \delta_a^{b'} \\ 
a_{11} = ( 1 + \delta_a^{b'} \delta_b^{a'} ) \delta_a^{a'} \delta_b^{b'}&
a_{12} = (1 + \delta_a^{a'} \delta_b^{b'} ) \delta_a^{b'} \delta_b^{a'}
\\ & 
\end{array} . 
\end{equation}
The coefficients $u_j$, for $j=2,\ldots,5$, are obtained from $u_1$ through 
appropriate place permutations of the upper indices ($\alpha, \beta, c', c'$) 
of the coefficient $M$ of Eq. (\ref{MUU*}). $u_2$ is obtained by the sum of 
the place permutations (12), (13), (14), (23), (24), (34), while $u_3$ by 
the sum of the permutations (123), (132), (124), (142), (134), (143), (234), 
(243); $u_4$ by permutations (12)(34), (13)(24), (14)(23), and finally $u_5$ 
by the place permutations (1234), (1243), (1324), (1342), (1423), (1432). The
results for $u_2$, $u_3$, $u_4$, $u_5$ are of the same form as Eq. (\ref{u1}) 
but with $a_k$ replaced by coefficients that we call $b_k$, $c_k$, $d_k$, 
$e_k$, respectively; they depend on sums of $a_k$'s. We will see below that 
not all them contribute to Eq. (\ref{Cqab(1)}); then, we show only the 
coefficients indexed by $k=11, 12$ that are important to that equation: 
\begin{eqnarray} \label{bcde-11-12} 
b_{11} & = & a_3 + a_5 + a_8 + a_9 + a_{11} + a_{12} \nonumber \\
b_{12} & = & a_2 + a_6 + a_7 + a_{10} + a_{11} + a_{12} \nonumber \\ 
c_{11} & = & a_2 + a_3 + a_5 + a_6 + a_7 + a_8 + a_9 + a_{10} \nonumber \\ 
c_{12} & = & c_{11} \nonumber \\ 
d_{11} & = & a_1 + a_4 + a_{12} \\ 
d_{12} & = & a_1 + a_4 + a_{11} \nonumber \\
e_{11} & = & a_1 + a_2 + a_4 + a_6 + a_7 + a_{10} \nonumber \\ 
e_{12} & = & a_1 + a_3 + a_4 + a_5 + a_8 + a_9 \nonumber \, . 
\end{eqnarray} 

For instance, the result for $J(\alpha,\beta,\gamma,\delta)$ can be written
as  
\begin{eqnarray} \label{M(a,b,g,d)} 
J(\alpha,\beta,\gamma,\delta) & = &
m_1 ( \delta_{\gamma}^{\alpha} \delta_{\delta}^{\beta} 
\delta_{\alpha'}^{c'} \delta_{\beta'}^{c'} ) + 
m_2 ( \delta_{\gamma}^{\alpha} \delta_{\delta}^{c'}
\delta_{\alpha'}^{\beta} \delta_{\beta'}^{c'} )  
\nonumber \\ & + &  
m_3 ( \delta_{\gamma}^{\alpha} \delta_{\delta}^{c'} 
\delta_{\alpha'}^{c'} \delta_{\beta'}^{\beta} ) + 
m_4 ( \delta_{\gamma}^{\beta} \delta_{\delta}^{\alpha} 
\delta_{\alpha'}^{c'} \delta_{\beta'}^{c'} ) 
\nonumber \\ & + & 
m_5 ( \delta_{\gamma}^{\beta} \delta_{\delta}^{c'}
\delta_{\alpha'}^{\alpha} \delta_{\beta'}^{c'} ) + 
m_6 ( \delta_{\gamma}^{\beta} \delta_{\delta}^{c'} 
\delta_{\alpha'}^{c'} \delta_{\beta'}^{\alpha} ) 
\nonumber \\ & + & 
m_7 ( \delta_{\gamma}^{c'} \delta_{\delta}^{\alpha}
\delta_{\alpha'}^{\beta} \delta_{\beta'}^{c'} ) + 
m_8 ( \delta_{\gamma}^{c'} \delta_{\delta}^{\alpha} 
\delta_{\alpha'}^{c'} \delta_{\beta'}^{\beta} ) 
\nonumber \\ & + & 
m_9 ( \delta_{\gamma}^{c'} \delta_{\delta}^{\beta} 
\delta_{\alpha'}^{\alpha} \delta_{\beta'}^{c'} ) +
m_{10} ( \delta_{\gamma}^{c'} \delta_{\delta}^{\beta}
\delta_{\alpha'}^{c'} \delta_{\beta'}^{\alpha} ) 
\nonumber \\ & + & 
m_{11} ( \delta_{\gamma}^{c'} \delta_{\delta}^{c'} 
\delta_{\alpha'}^{\alpha} \delta_{\beta'}^{\beta} ) + 
m_{12} ( \delta_{\gamma}^{c'} \delta_{\delta}^{c'}
\delta_{\alpha'}^{\beta} \delta_{\beta'}^{\alpha} ) , \nonumber \\
\end{eqnarray}
where 
\begin{equation} \label{mjs} 
m_k = Aa_k + Bb_k + Cc_k + Dd_k + Ee_k , \quad k=1,\ldots,12 . 
\end{equation}
From Eq. (\ref{M(a,b,g,d)}) we construct the coefficients $J(\alpha,\beta,c,c)$
and $J(c,c,\alpha,\beta)$, take the difference of them and sum with respect to $c$,
$c'$. The result is given by Eq. (\ref{M2-M1}), where
\begin{eqnarray}
n_1 & = & m_{11} + m_{12} \label {n1} \\
n_2 & = & m_2 - m_3 - m_9 + m_{10} - Nm_{11} \label{n2} \\
n_3 & = & m_2 - m_3 - m_9 + m_{10} + Nm_{12} \label{n3} \, .
\end{eqnarray}
From Eqs. (\ref{n2}) and (\ref{n3}) we see that
\begin{equation}\label{n3-n2}
n_3-n_2=Nn_1
\end{equation}

Eqs. (\ref{Cqab(1)}) and (\ref{M2-M1}) leads to Eq. (\ref{Cqab(1)-2}), the result
being dependent on $n_1$, and $n_2$, $n_3$ through the difference $n_3-n_2=Nn_1$.
From Eqs. (\ref{bcde-11-12}), (\ref{mjs}) and (\ref{n1}), $n_1$ is given by
\begin{eqnarray} \label{n1-2}
n_1 & = & ( A + 2B + D ) ( a_{11} + a_{12} ) + 2 (D+E) ( a_1 + a_4 )
\nonumber \\ & + &
( B + 2C + E ) (a_2 + a_3 + a_5
\nonumber \\  & + & a_6 + a_7 + a_8 + a_9 + a_{10} ) .
\end{eqnarray}
Finally, Eqs. (\ref{ABCDE}) and (\ref{ajs}) give
\begin{eqnarray} \label{n1-3}
 n_1 & = & \frac{1}
{N^2 ( N^2-1 ) ( N+2 ) ( N+3 ) }
\left\{
2 ( 1 + \delta_a^b ) ( 1 + \delta_{a'}^{b'} )
\right.
\nonumber \\ & + &
\left( N+1 \right) \left( N+2 \right)
\left( \delta_a^{a'} \delta_b^{b'} +
\delta_a^{b'} \delta_b^{a'} \right)^2
\nonumber \\ & - &
\left( N+1 \right) \left[
\delta_b^{a'} + \delta_b^{b'} + \delta_a^{a'} + \delta_a^{b'} \right.
\nonumber \\ & & \quad +
2 \, \delta_a^b \delta_{a'}^{b'} \left( \delta_b^{b'} \delta_{a'}^a +
\delta_b^{a'} \delta_{b'}^a \right) +
2 \left( \delta_b^{b'} \delta_{a'}^b \delta_{b'}^{a'} \right.
\nonumber \\ & & \quad + \left. \left. \left.
\delta_a^b \delta_b^{a'} \delta_{a'}^a +
\delta_a^b \delta_b^{b'} \delta_{b'}^a +
\delta_a^{b'} \delta_{a'}^a \delta_{b'}^{a'} \right)
\right] \right\}
\end{eqnarray}



\end{document}